\input harvmac.tex
\vskip 2in
\Title{\vbox{\baselineskip12pt
\hbox to \hsize{\hfill}
\hbox to \hsize{\hfill IC/98/224}}}
{\vbox{\centerline{On the NSR formulation of String Theory on 
$AdS_5\times{S^5}$}
\vskip 0.3in
{\vbox{\centerline{}}}}}
\centerline{Dimitri Polyakov\footnote{$^\dagger$}
{polyakov@ictp.trieste.it}}
\medskip
\centerline{\it The Abdus Salam International Centre for Theoretical Physics}
\centerline{\it Strada Costiera,11 }
\centerline{\it I-34014  Trieste, ITALIA}
\vskip .5in
\centerline {\bf Abstract}
We discuss the NSR formulation of the superstring action
on $AdS_5\times{S^5}$ proposed recently by Kallosh and
Tseytlin in the Green-Schwarz formalism.We show that 
the stress-energy tensor corresponding to
the NSR action for AdS superstring contains 
the branelike terms, corresponding
to exotic massless vertex operators (refered to as the branelike
vertices)- the 5-form $e^{\phi}\psi_{m1}...\psi_{m5}$
and the 3-form $\partial(e^{\phi}\psi_{m_1}..\psi_{m_3})$,
multiplied by  $\partial{X_m}$.
The corresponding sigma-model action has the manifest
$SO(1,3)\times{SO(6)}$ invariance of
superstring theory on $AdS_5\times{S^5}$.
We argue that adding the branelike terms is equivalent to
curving the space-time to obtain the $AdS_5\times{S^5}$ background.
We commence the study of the proposed NSR sigma-model
by analyzing the scattering amplitudes involving the branelike
vertex operators.The analysis shows quite an unusual momentum
dependence of these scattering amplitudes.

{\bf Keywords:} 
{\bf PACS:}$04.50.+h$;$11.25.Mj$. 
\Date{December 98}
\vfill\eject
\lref\wil{K.G.Wilson, Phys.Rev.D10:2445-2459,1974}
\lref\kts{R.Kallosh, A.Tseytlin,{\bf J.High Energy Phys.9810:016,1998}}
\lref\pe{I.Pesando,{\bf JHEP11(1998)002}}
\lref\mets{R.Metsaev, A.Tseytlin,{\bf Phys.Lett.B436:281-288,1998}}
\lref\ampf{S.Gubser,I.Klebanov, A.M.Polyakov, 
{\bf Phys.Lett.B428:105-114}}
\lref\amps{A.M.Polyakov{\bf Nucl.Phys.B486:23-33,1997}}
\lref\ampt{A.M.Polyakov,hep-th/9809057}
\lref\wit{E.Witten{\bf Adv.Theor.Math.Phys.2:253-291,1998}}
\lref\gibbons{G.Gibbons,M.Green,M.Perry
{\bf Phys.Lett. B370 (1996)}}
\lref\gibb{G.Gibbons, P.Townsend{\bf Phys.Rev.Lett.71(1993),5223}}
\lref\malda{J.Maldacena,hep-th/9711200}
\lref\kleb{S.Gubser,A.Hashimoto,I.Klebanov,J.Maldacena,
{\bf Nucl.Phys.B472:231-248,1996}}
\lref\myself{D.Polyakov{\bf Phys.Rev.D57:2564-2570,1998}}
\lref\self{D.Polyakov,{\bf Nucl.Phys.B468:155-162,1996}}
\lref\gs{M.Green,J.Schwarz,{\bf Phys.Lett.B136 (1984) 367}}
\lref\sw{J.Schwarz{\bf Nucl.Phys.B226(1983) 269}}
\lref\krr{R.Kallosh,J.Rahmfeld, A.Rajaraman,hep-th/9805217}
\lref\mts{R.Metsaev,A.Tseytlin,hep-th/9806095}
\lref\polchinski{J.Polchinski,{\bf Phys.Rev.Lett.75:4724-4727,1995}}
\lref\ramf{R.Kallosh, J.Rahmfeld, hep-th/9808038}
\centerline{\bf Introduction}  
Recently the classical type IIB Green-Schwarz (GS)
~\refs{\gs}
superstring action 
 in the maximally supersymmetric
$AdS_5\times{S^5}$ background ~\refs{\sw,\mts,\krr}
has been constructed
 by Metsaev and Tseytlin ~\refs{\mets} and later simplified 
in the work by Kallosh and Tseytlin by fixing the
$\kappa$-symmetry ~\refs{\kts}; see also ~\refs{\pe,\ramf}.
In accordance with the conjecture that relates the  dynamics of
$ N=4 D=4$ super Yang-Mills theory  with the type IIB superstring
theory and supergravity ~\refs{\ampf,\amps,\ampt,\wit,\malda}
in the AdS background, string theory
on the AdS space appears to be a plausible
candidate for the string-theoretic description of the large N limit
of super Yang-Mills theory. On the condition that the 
analogue of the Maldacena's conjecture is formulated for the 
non-supersymmetric case and a suitable mechanism is found for supersymmetry
breaking, studying the type IIB  string dynamics in  the
$AdS_5\times{S^5}$ may be relevant for the understanding of the 
problem of confinement from the string-theoretic point of view.
Some time ago, in an originally independent development
we have noticed the existence of certain new exotic massless states
in the spectrum of NSR string theory which apparently
contain information about non-perturbative aspects of string
dynamics.These open string states are described by BRST-invariant
two-form and five-form vertex operators of 
essentially nonzero ghost numbers.
They do not correspond to
any known massless excitations in perturbative spectrum of an open 
string; while  precise physical interpretation of these 
vertex operators is
still largely a puzzle, in our previous works ~\refs{\myself} we have
 made a preliminary conjecture
(though unfortunately still vague and incomplete at the time)
  that these excitations 
(which we refered to as branelike states since the superalgebraic
arguments point at their connection with M-brane dynamics) 
may be relevant
to the dynamics of non-Abelian gauge theories.
In this paper we shall discuss the connection between
these two developments; namely we shall attempt to present the
NSR formulation of the Kallosh-Metsaev-Tseytlin type action of a type
IIB string on  AdS; we shall argue that the NSR version
of stress-energy tensor corresponding to
 the superstring action on $AdS_5\times{S^5}$ is related to the one
of
a sigma-model with the  background determined by above mentioned
branelike excitations.
The paper is organized as follows.
In the first section we review the branelike states and
give  physical arguments for their appearance as a result
of the internal normal ordering in Ramond-Ramond vertices;
then, by using the open string description of D3-branes,
 we will perform the GS-NSR transformation for
the GS stress-energy tensor in the
presence of D3-branes and observe the
appearance of the terms related to the branelike states.

The resulting NSR stress-energy tensor
(as well as the corresponding action)
 will have the manifest $SO(1,3)\times{SO(6)}$
symmetry of superstring theory on
curved $AdS_5\times{S^5}$ background
In the second section we compute and study the properties of
scattering amplitudes that involve branelike states.
More precisely, we study the influence of the branelike states
on the gravitational lensing of the Ramond-Ramond states by D p-branes
and observe that at  certain values of the momentum
the presence of these exotic states leads to additional simple poles
in scattering amplitudes.

\centerline{\bf 1.The type IIB 
action on $AdS_5\times{S^5}$ in the NSR formalism and branelike states}

 Consider   the  Ramond-Ramond vertex operators at zero momentum 
in $(-1/2,-1/2)$ and $(-3/2,-3/2)$-pictures
 on a disc:
\eqn\grav{\eqalign
{V^{(-1/2,-1/2)}(k,z,\bar{z})
=e^{-{1\over2}\phi}\Sigma^\alpha(z)
e^{-{1\over2}{\bar\phi}}\bar\Sigma^\beta(\bar{z})e^{ikX}(z,\bar{z})
{\Gamma^{m_1...m_q}_{\alpha\beta}}F_{m_1...m_q}(k)\cr
V^{(-3/2,-3/2)}(k,z,\bar{z})
\cr=
(e^{-{3\over2}\phi}+{1\over2}
e^{\chi-{5\over2}\phi}\partial^2{c})\Sigma^\alpha(z)
(e^{-{3\over2}{\bar\phi}}+{1\over2}e^{\bar\chi-{5\over2}\bar\phi}
\bar\partial^2{\bar{c}})\cr\times
\bar\Sigma^\beta(\bar{z})e^{ikX}(z,\bar{z}){\Gamma^{m_1...m_q}_{\alpha\beta}}
F_{m_1...m_q}(k)}}

where $\phi(z)$ and $\chi(z)$ are free fields that appear in the
bosonization of the NSR superconformal ghosts $\beta$,$\gamma$;
$\Sigma^\alpha$ is  spin operator for  NSR matter fields.
In the future, unless stated otherwise, we will be dropping
the terms depending on fermionic ghosts since
these terms will be insignificant for our computations of correlation
functions (though in principle these ghost terms are necessary
to insure the BRST invariance of vertex operators)
If a Ramond-Ramond vertex is placed on a disc 
and the boundary is present, the holomorphic and anti-holomorphic
matter and ghost spin operators are no longer independent but
they are related as:
\eqn\grav{\eqalign{\bar\phi(\bar{z})=\phi(\bar{z}),
\bar\chi(\bar{z})=\chi(\bar{z})\cr
\bar\Sigma^\alpha(\bar{z})={M^{(p)\alpha}_\beta}\Sigma^\beta(\bar{z})\cr
M_{\alpha\beta}\equiv({\Gamma^0...\Gamma^p})_{\alpha\beta}}}
The expression for the matrix $M^{(p)}_{\alpha\beta}$
implies that the Dirichlet boundary conditions are
imposed on p out of 10 $X^m$'s while the Neumann conditions
are imposed on the rest.
As long as  the vertices (1) are far from the edge of a D-brane
(that is, $z\neq{\bar{z}}$) one may neglect the interaction
 between holomorphic and anti-holomorphic spin operators;
however, as one approaches the boundary of the disc where $z=\bar{z}$
the internal normal ordering must be performed inside the 
Ramond-Ramond vertex operators in order to remove the singularities
that arise in the O.P.E. between the spin operators located
at $z$ and $\bar{z}$. 
Adopting the notation
${\not{F}}^{(q)}(k)\equiv{\Gamma^{m_1}...\Gamma^{m_q}}F_{m_1...m_q}(k)$
we find that the result of the normal ordering is given by:
\eqn\grav{\eqalign{lim_{{z,\bar{z}}\rightarrow{s}}
:e^{-{3\over2}\phi}\Sigma_\alpha:
(z){{\not{F}}^{(q)}_{\alpha\beta}}:e^{-{3\over2}\bar\phi}\bar\Sigma_\beta
(\bar{z}):\cr\sim{1\over{z-\bar{z}}}Tr({{\not{F}}^{(q)}}M^{(p)}
\Gamma^{m_1...m_5})e^{-3\phi}\psi_{m_1}...\psi_{m_5}(s)+...\cr
{lim_{{z,\bar{z}}\rightarrow{s}}}
:e^{-{3\over2}\phi}\Sigma_\alpha:
(z){{\not{F}}^{(q)}_{\alpha\beta}}:e^{-{1\over2}\bar\phi}\bar\Sigma_\beta
(\bar{z}):\cr\sim{1\over{z-\bar{z}}}Tr({{\not{F}}^{(q)}}M^{(p)}
\Gamma^{m_1m_2})e^{-2\phi}\psi_{m_1}\psi_{m_2}(s)+...}}
where we have dropped the less singular terms in the O.P.E.
as well as full derivatives.
We see that due to the internal normal ordering at the
boundary of the disc the
Ramond-Ramond vertex operators degenerate into massless
 $open-string$ vertices - the two-form 
$Z_{mn}=e^{-2\phi}\psi_m\psi_n$ and the five-form
$Z_{m_1...m_5}=e^{-3\phi}\psi_{m_1}...\psi_{m_5}$.
As for the five-form state, there also exists the ``dual''
version of it in the $+1$-picture:
$Z_{m_1...m_5}^{(+1)}\equiv:\Gamma^4{Z_{m_1...m_5}}:=
{e^\phi}\psi_{m_1}...\psi_{m_5}$ where $\Gamma^4$ is
the normally ordered fourth power of picture-changing operator.
Actually it is this picture-changed version of the 5-form 
(i.e. the one in the $+1$-picture) that
will appear in the NSR formulation of the AdS superstring action.

As it is the well-known fact
that there are no two-form and five-form particles
in the $peturbative$ spectrum of an open string,
giving a proper physical interpretation to these
$new$ massless states in the spectrum of a superstring
is a challenging puzzle. In our previous works
~\refs{\myself} we have shown that two-form and five-form
vertices (3)  appear as central terms in the 
space-time superalgebra for NSR superstring theory when the
supercharges are taken in non-canonical pictures.
Since p-form central terms in a SUSY algebra are always related to 
the presence of p-branes
we have  argued that the operation of picture-changing
(applied to space-time supercharges) is a worldsheet
interpretation of the S-duality transformation and the 
open string vertex
operators (3) have essentially non-perturbative origin and 
represent the M-branes, though the precise relation of
these vertex operators to the dynamics of branes is still
 unclear.
In the current paper we shall discuss the role that
 these new superstring states may play in the recently proposed
AdS-CFT correspondence ~\refs{\malda,\wit,\ampf,\amps,\ampt}
and in building the string-theoretic
approach to Yang-Mills theories.
In the light of the newly found relation between the 
Yang-Mills correlation functions  and the minimum of the 
supergravity action on the AdS space, the type IIB
superstring theory on AdS (which has the AdS supergravity
its low-energy limit) appears a plausible candidate for
the ansatz to describe the dynamics of large N gauge theories.
It is therefore crucial to understand the dynamics  of
string theory in the AdS background in order to 
 deepen our understanding of Yang-Mills dynamics and 
CFT in various dimensions.
In this paper we shall claim that  
in terms of NSR formalism superstring
theory on the  AdS space ( presumably describing
the large N limit of gauge theories) corresponds to
a sigma model with $flat$ space-time metric
and with
 massless background fields corresponding to
the branelike vertex operators (2).
There are two dual ways of regarding the superstring dynamics in the
field of D-branes.The first way is to consider the closed GS superstring
propagating in the AdS background corresponding to the 
near-horizon geometry of D3-brane solution.
The $AdS_5\times{S^5}$ action in the GS formalism is given by
~\refs{\kts}:

\eqn\grav{\eqalign{S={-{1\over2}}\int{d^2}z\lbrace
{\sqrt{-g}}g^{ij}(y^2(\partial_i{x^p}-2i{\bar\theta}{\Gamma^p}\partial_i
{\theta})(\partial_j{x^p}-2i\bar\theta\Gamma^p\partial_j\theta
 )+{1\over{y^2}}{\partial_i}y^t\partial_j{y^t})\cr
+4\epsilon^{ij} \partial_i{y^t}\bar\theta{\Gamma^t}
\partial_j\theta\rbrace }}

where $X^p,p=0...3$ and $Y^t, t=4...10$  are the longitudinal
and transverse coordinates with respect to the worldvolume of
the D3-branes.
The second way to explore the GS superstring dynamics in the
D-brane background (which is equivalent up to the D-brane's massive modes)
is to study the propagation of the GS closed supersting
in the $flat$ space-time in the presence of open Dirichlet strings
~\refs{\polchinski,\kleb}.In the presence of open strings
the holomorphic and antiholomorphic fields $\theta$ and
$\bar\theta$ are no longer independent.
The GS superstring action in the flat space-time is given by:
\eqn\grav{\eqalign{-{1\over2}\int{d^2{z}}\lbrace
(\partial{X^m}-i\bar\theta\Gamma^m\partial\theta)
(\bar\partial{X_m}-i\bar\theta^\alpha
(\Gamma_{\alpha\beta})_m\bar\partial\theta^\beta)\cr
-2i\epsilon^{ij}\partial_i{X^m}\bar\theta^\alpha
{(\Gamma_{\alpha\beta})_m}\partial_j\theta^\beta\rbrace
}} 

The corresponding stress-energy tensor is given by
\eqn\grav{\eqalign{T_{ij}=\Pi^m_i\Pi_{mj}\cr
\Pi^m_i=\partial_i{X^m}+i\bar\theta^\alpha\Gamma^m_{\alpha\beta}
\partial_i\theta^\beta}}
In the
flat space-time one may use the standard relations
that exist (up to picture-changing transformation)
between GS fermionic variable $\theta$ and NSR spin operators
in ten-dimensional 
space-time:$\theta^\alpha={e^{{\phi\over2}}}\Sigma^\alpha+ghosts$.

In the presence of the D3-branes the space-time fermionic
fields $\theta$ and $\bar{\theta}$ are not independent but
 are related to each other according to (2).
For the D3-branes this relation is given by:
\eqn\lowen{\bar\theta_\alpha(\bar{z})=
\lbrack{
i{(\Gamma^0...\Gamma^3)}_{\alpha\beta}\rbrack
\theta_\beta(\bar{z})}}
Using (7) let us now proceed with the formulation of the
NSR  analogue of the stress-energy tensor (6).
Because of (7) the normal reordering between $\bar\theta$ and
$\theta$ must be made , similar to the one performed in
the Ramond-Ramond vertex operators in (3). 
Using (7) and
 the formula for the O.P.E. between two spin operators:
\eqn\lowen{:\Sigma_\alpha(z)::\Sigma_\beta(w):\sim
{{\epsilon_{\alpha\beta}}\over{{(z-w)}^{5\over4}}}+
{\sum_p}{{\Gamma^{m_1...m_p}_{\alpha\beta}
\psi_{m_1}...\psi_{m_5}}\over{(z-w)^{{5\over4}-p}}}}

one finds that the stress-energy
tensor $T_{zz}$ of
 (6) rewritten in the NSR
formalism is given by:
\eqn\grav{\eqalign{{T_{{zz}{NSR}}}= 
{1\over2}\lbrack\partial{{X_m}}\partial{{X^m}}
+\lbrace
e^\phi(\partial\phi
\partial\psi_\mu\partial{X^\mu}+\psi_m\partial\psi^m\psi_n
\partial{X^n})\cr+
\epsilon^{p_1p_2p_3p_4}
{\lbrack}e^\phi\psi_{p_1}\psi_{p_2}\psi_{p_3}\psi_{p_4}\psi^t\partial{y^t}
+\partial(e^\phi\psi_{p_1}
\psi_{p_2}\psi_{p_3})\partial{{\tilde{x}}_{p_4}}\rbrack
\cr+{1\over2}:\Gamma\Gamma\psi\partial\psi:+ghosts
\rbrack
 }}
Here $\epsilon^{p_1...p_4}$ is the rank 4 antisymmetric tensor
for the D3 brane vorldvolume ( the indices $p_1,...p_4$ run from 0 to 3,
 $t$ runs from 4 to 9);$m=(p,t)$ and $n$ are ten-dimensional
indices,$X^m\equiv({\tilde{x^p}},y^t)$.
 We have chosen the conformal gauge
for the worldsheet metric and dropped full derivative terms.
The tensor (9) may be rewritten in the form where the kinetic term
for the NSR fermions
 (quadratic in $\psi$) appears in a
more transparent way:
\eqn\grav{\eqalign{T_{{zz}{NSR}}=
\lbrack{1\over2}\partial{X^m}\partial{X_m}
+\lbrace{1\over2}
:(\Gamma+\Gamma\Gamma)\psi^{m}\partial\psi_{m}:\cr+
\epsilon^{p_1p_2p_3p_4}\lbrack{e^\phi}\psi_{p_1}\psi_{p_2}
\psi_{p_3}\psi_{p_4}\psi_t\partial{y^t}+\partial(e^\phi\psi_{p_1}
\psi_{p_2}\psi_{p_3})\partial{{\tilde{x}}_{p_4}}\rbrack+ghosts
\rbrack}}
Here $\Gamma\equiv:\delta(\gamma)(G_{matter}+G_{ghost}):$ 
 is the
picture-changing operator where $G$ is the matter $+$ ghost
supercurrent.
Up to the picture-changing
transformation,
the kinetic part of the resulting NSR stress-energy tensor is
the same as in the
NSR tensor in flat ten-dimensional space-time without branes.
 The stress-energy tensor (10) corresponds to 
 a sigma-model with flat space-time metric and the
branelike part; it is the branelike states that break
the SO(1,9) Lorentz symmetry of the flat space-time action
to $SO(1,3)\times{SO(6)}$ of that on $AdS_5\times{S^5}$.
The branelike part corresponds to the term
$\sim\bar\theta^\alpha\Gamma^m_{\alpha\beta}
\partial\theta^\beta\partial{X_m}$ in the stress-energy tensor.
It is important to note that it is crucial to have the 
D-3 branes (that give rise to the
curved ${AdS_5\times{S^5}}$ background in the ``dual'' approach) 
in order to include the 
branelike states (3) in the sigma-model; in the absence of D3-branes
(that is, in the flat space-time)
there would be no $\Gamma^0...\Gamma^3$ factor in (7) and
the 5-form and the 3-form terms in (9), (10) would have vanished.
For instance, in the absence of the above gamma-matrix factor
the corresponding 5-form term would be given by
$\sim{Tr}(\Gamma^{\mu_1...\mu_5}\Gamma^\nu)e^\phi
\psi_{\mu_1}...\psi_{\mu_5}\partial{X_\nu}$
but this expression obviously vanishes since , due to anticommutation
relations between the gamma-matrices, one would inevitably have to
contract a pair of anticommuting fermions
$\psi_\mu$,$\psi_\nu$ with the Minkowski metric tensor $\eta^{\mu\nu}$.
 In the next section of this paper we will explore the properties
of the sigma-model corresponding to the energy-momentum tensor (10) 
and study the correlation functions of the branelike vertex operators.
The calculation will show a rather peculiar momentum dependence of
these correlators.

\centerline{\bf Correlation Functions of Branelike States}
Rather than computing the amplitudes involving
the open string branelike states (3)
in a straightforward way we will choose, for the sake of simplicity,
to study the problem of gravitational lensing of the
 the Ramond-Ramond
particles in the non-standard $(-3/2,-3/2)$-picture in the D-brane field
 which in some way is equivalent to studying
the properties of branelike states (see the comment below).
In case of the RR particles in the $standard$ $(-1/2,-1/2)$ picture
the appropriate scattering amplitude has been computed in ~\refs{\kleb}
The essential difference between these two computations is related
to the fact that the Ramond-Ramond vertex $V^{(-3/2,-3/2)}$ of (1)
(unlike the RR vertex in the standard picture)
evolves into the exotic five-form state (3) on the boundary of the disc,
as we have shown above.Therefore one should expect that
the scattering amplitude for the RR states in the D-brane field
is $picture-dependent$.
The computation will show that in comparison with the
standard scattering amplitude for the $(-1/2,-1/2)$ RR particles
there are additional simple poles that
originate from integrating over the region $r\sim{1}$ 
(i.e.near the boundary
of the disc) where the $(-3/2,-3/2)$ RR vertex operator (2) becomes
the branelike 5-form state (3), therefore 
these extra poles in the scattering amplitude
are apparently of the branelike origin.
Consider the correlation function:
\eqn\grav{\eqalign{W(z,\bar{z},w,\bar{w},k_1,k_2)
=\langle e^{-{3\over2}\phi}\Sigma_{\alpha_1}(z)e^{-{3\over2}\bar\phi}
\bar\Sigma_{\beta_1}(\bar{z}){{\not{F}}^{(p)}_{\alpha_1\beta_1}}
e^{ik_1X}(z,\bar{z})\cr\times
e^{\phi\over2}\Sigma_{\gamma_2}\Gamma^m_{\gamma_2\alpha_2}
\partial{X_m}(w)e^{{\bar\phi}\over2}\bar\Sigma_{\delta_2}
{\Gamma^n_{\delta_2\beta_2}}\bar\partial{X_n}
{{\not{F}}^{(p)}_{\alpha_2\beta_2}}(k_2)e^{ik_2{X}}(w,\bar{w})\rangle\cr=
(F^{(p)}(k_1)M^{(q)})_{\alpha_1\rho_1}
(\Gamma^mF^{(p)}(k_2)\Gamma^nM^{(q)})_{\gamma_2\rho_2}
\langle e^{-{{3\phi}\over2}}(z)e^{-{{3\phi}\over2}}(\bar{z})e^{\phi\over2}(w)
e^{\phi\over2}(\bar{w})\rangle\cr\times
\langle\Sigma_{\alpha_1}(z)\Sigma_{\rho_1}(\bar{z})\Sigma_{\gamma_2}(w)
\Sigma_{\rho_2}(\bar{w})\rangle\cr\times
\langle e^{ik_1{X}}(z)e^{ik_1{X^\pm}}(\bar{z})\partial{X_m}e^{ik_2{X}}(w)
f(n)\partial{X_n}e^{ik_2{X^\pm}}(\bar{w})\rangle
}}
Here $f(n)=-1$ for the longitudinal indices (with respect to the D3 brane
worldvolume) and 1 for the transverse ones,$X^\pm\equiv{f(n)}X$
To compute the $X$-independent
part $W_0$ of this correlation function (that involves ghosts and
fermions)
one notes that
\eqn\lowen{W_0\sim{{\langle e^{-{{3\phi}\over2}}
e^{-{{3\phi}\over2}}e^{{\phi\over2}}
e^{{\phi\over2}}\rangle}\over{\langle e^{-{\phi\over2}}
e^{-{\phi\over2}}e^{-{\phi\over2}}
e^{-{\phi\over2}}\rangle}}\langle\Lambda\Lambda\Lambda\Lambda\rangle}
where $\Lambda$ are the full (ghost + matter) (1,0) spin operators
in the standard $-{1\over2}$-picture.

A bit long but elementary computation  gives the following result
for the zero momentum part of the correlator:
\eqn\grav{\eqalign{W(z,\bar{z},w,\bar{w},k_{1,2}=0)=
Tr(\Gamma^r{{\not{F}}^{(p)}}(k_1)M^{(q)})Tr(\Gamma^r\Gamma^m
{{\not{F}}^{(p)}}(k_2)\Gamma^n{M^{(q)}})
{{|z-{\bar{w}}|^2}\over{(w-\bar{w})(z-\bar{z})^3}}\cr+
8Tr({\not{F}}^{(p)}(k_1)M^{(q)}\Gamma^m{\not{F}}^{(p)}(k_2)\Gamma^n
M^{(q)}){1\over{{(z-{\bar{z}})}^2}}}}
The kinetic part of the correlator is given by:
\eqn\grav{\eqalign{\langle e^{ik_1{X}}(z)e^{ik_1{X^\pm}}
(\bar{z})\partial{X_m}e^{ik_2{X}}(w)
f(n)\partial{X_n}e^{ik_2{X^\pm}}(\bar{w})\rangle\cr=
\delta(k_1^{||}+k_2^{||})\lbrack
(z-\bar{z})^{\lbrace{k_1^2}\rbrace}(w-\bar{w})^{\lbrace{k_2^2}\rbrace}
|z-\bar{w}|^{2\lbrace{k_1k_2}\rbrace}|z-w|^{2(k_1k_2)}\rbrack\cr\times
\lbrack{{\delta_{mn}f(n)}\over{(w-\bar{w})^2}}+
({{{k_{2n}}f(n)}\over{w-\bar{w}}}+k_{1n}f(n)({1\over{z-\bar{w}}}-
{1\over{\bar{z}-w}})+k_{1n}({1\over{z-w}}-{1\over{\bar{z}-\bar{w}}}))\cr\times
({{{k_{2m}}f(m)}\over{w-\bar{w}}}+k_{1m}f(m)({1\over{z-\bar{w}}}-
{1\over{\bar{z}-w}})+k_{1m}({1\over{z-w}}-{1\over{\bar{z}-\bar{w}}}))\cr+
{{\delta_{mn}}\over{w-\bar{w}}}(2{\lbrace{k_1k_2}\rbrace}(
{1\over{z-\bar{w}}}-{1\over{\bar{z}-w}})+2(k_1k_2)({1\over{\bar{z}-\bar{w}}}
-{1\over{z-w}}))\rbrack}}
where we  defined
 $\lbrace{k_1,k_2}\rbrace\equiv 
k_{1p}k_2^p-k_{1t}k_2^t\equiv(k_1k_2)^{||}-(k_1k_2)^{\perp}$
and similarly for $\lbrace{k_{1,2}^2}\rbrace$;
 $(k_1k_2)$  stands for the regular scalar product
and $k^{||}\equiv{k_pk^p}$;$p=0,...q;t=q+1,...9$.
 
So far we were considering the correlation functions on a half-plane.
Let us make the conformal transformation 
$(z,\bar{z})\rightarrow(u,\bar{u})$ from the half-plane
 to the disc:
\eqn\lowen{z={i\over2}({{u+i}\over{u-i}})}
Let us furthermore fix $w={-{i\over2}}
$ so that its image is at the center of
the disc.In terms of the disc coordinates
 $u\equiv{r}e^{i\varphi}$ the correlation function (14) becomes:
\eqn\grav{\eqalign{W(r,\varphi,k_1,k_2)=
(r^2+1-2rsin\varphi)^{-\lbrace{k_1}^2\rbrace-\lbrace{k_1k_2}\rbrace
-(k_1k_2)}(r^2-1)^{2(k_1^{||})^2}r^{2(k_1k_2)}
\cr\times\lbrack{Tr}(\Gamma^r{\not{F(k_1)}}^{(p)}
M^{(q)})Tr(\Gamma^r\Gamma^m{\not{F(k_2)}}^{(p)}\Gamma^n{M^{(q)}})
{{(r^2+1-2rsin\varphi)^2}\over{(r^2-1)^3}}\cr+
8Tr({\not{F}}^{(p)}(k_1)M^{(q)}\Gamma^m{\not{F}}^{(p)}(k_2)\Gamma^n
M^{(q)}){{(r^2+1-2rsin\varphi)^2}\over{(r^2-1)^2}}\rbrack\cr\times
\lbrack{\delta_{mn}}f(n)+(k_{2n}f(n)+4k_{1n}rcos\varphi+k_{1n}
(1+{{sin\varphi}\over{r}}))\cr\times(k_{2m}f(m)+4k_{1m}rcos\varphi+k_{1m}
(1+{{sin\varphi}\over{r}}))\cr+\delta_{mn}(8\lbrace{k_1k_2}rcos\varphi+
2(k_1k_2)(1+{{sin\varphi}\over{r}}))\rbrack}}
Since the momentum in the longitudinal directions
(with respect to the D3-brane worldvolume) is conserved
one has 
${\lbrace{k_1}^2\rbrace+\lbrace{k_1k_2}\rbrace
+(k_1k_2)}=0$.
Integrating over $r$ and $\varphi$ we obtain the following expression
for the amplitude:
\eqn\grav{\eqalign{A(k_1,k_2)\equiv{\int_0^1}rdr{\int_0^{2\pi}}d\varphi
W(r,\varphi,k_1,k_2)\cr=
4k_{1n}k_{1m}B(4+(k_1k_2),2{(k_1^{||})^2}-2)
{Tr}(\Gamma^r{\not{F(k_1)}}^{(p)}
M^{(q)})Tr(\Gamma^r\Gamma^m{\not{F(k_2)}}^{(p)}\Gamma^n{M^{(q)}})
\cr+8Tr({\not{F}}^{(p)}(k_1)M^{(q)}\Gamma^m{\not{F}}^{(p)}(k_2)\Gamma^n
M^{(q)})B(4+(k_1k_2),2{(k_1^{||})^2}-1)\cr+
({Tr}(\Gamma^r{\not{F(k_1)}}^{(p)}
M^{(q)})Tr(\Gamma^r\Gamma^m{\not{F(k_2)}}^{(p)}\Gamma^n{M^{(q)}})
B(3+(k_1k_2),2{(k_1^{||})^2}-2)\cr+
8Tr({\not{F}}^{(p)}(k_1)M^{(q)}\Gamma^m{\not{F}}^{(p)}(k_2)\Gamma^n
M^{(q)})B(3+(k_1k_2),2{(k_1^{||})^2}-1))\cr\times
(5k_{1m}k_{1m}+(2(k_1k_2)+f(n))\delta_{mn}+
k_{2n}k_{2m}f(n))\cr+
({Tr}(\Gamma^r{\not{F(k_1)}}^{(p)}
M^{(q)})Tr(\Gamma^r\Gamma^m{\not{F(k_2)}}^{(p)}\Gamma^n{M^{(q)}})
B(2+(k_1k_2),2{(k_1^{||})^2}-2)\cr+
8Tr({\not{F}}^{(p)}(k_1)M^{(q)}\Gamma^m{\not{F}}^{(p)}(k_2)\Gamma^n
M^{(q)})B(2+(k_1k_2),2{(k_1^{||})^2}-1))\cr\times
({7\over2}k_{1m}k_{1n}+(4(k_1k_2)+3f(n))\delta_{mn}\cr+
3k_{2n}k_{2m}f(m)f(n)-k_{1m}k_{2n}f(n)-k_{2m}k_{1n}f(m))\cr+
({Tr}(\Gamma^r{\not{F(k_1)}}^{(p)}
M^{(q)})Tr(\Gamma^r\Gamma^m{\not{F(k_2)}}^{(p)}\Gamma^n{M^{(q)}})
B(1+(k_1k_2),2{(k_1^{||})^2}-2)\cr+
8Tr({\not{F}}^{(p)}(k_1)M^{(q)}\Gamma^m{\not{F}}^{(p)}(k_2)\Gamma^n
M^{(q)})B(1+(k_1k_2),2{(k_1^{||})^2}-1))\cr\times
({3\over2}k_{1m}k_{1n}+\delta_{mn}f(n)+k_{2m}k_{2n}f(m)f(n)-
k_{1n}k_{2m}f(n)-k_{1m}k_{2n}f(n))\cr+
{1\over2}({Tr}(\Gamma^r{\not{F(k_1)}}^{(p)}
M^{(q)})Tr(\Gamma^r\Gamma^m{\not{F(k_2)}}^{(p)}\Gamma^n{M^{(q)}})
B((k_1k_2),2{(k_1^{||})^2}-2)\cr+
8Tr({\not{F}}^{(p)}(k_1)M^{(q)}\Gamma^m{\not{F}}^{(p)}(k_2)\Gamma^n
M^{(q)})B((k_1k_2),2{(k_1^{||})^2}-1))k_{1m}k_{1n}}}
where $B(p,q)={{\Gamma(p)\Gamma(q)}\over{\Gamma(p+q)}}$ is the
Euler beta function.
Some comments should be made about the momentum behaviour 
of the
amplitude (17).First of all it is noteworthy that,
in addition to the poles present in the ``standard'' scattering
amplitude of ~\refs{\kleb} 
there are additional physical singularities (i.e. the simple poles)
in the amplitude (17) at $(k_1^{||})^2=0,{1\over{{2}}},1$ that
originate  from  integration 
over the region $r\sim{1}$, i.e.near the boundary of the disc where,
as we pointed out earlier, the Ramond-Ramond vertex in the
$(-3/2,-3/2)$-picture becomes the 5-form branelike state (3).
One may therefore conclude that it is exclusively the $boundary$
5-form state (3) that gives rise to the additional 
simple poles in the scattering amplitude
(17). Note that these poles did not arise in the calculation
in ~\refs{\kleb}
 as the RR vertex in the $(-1/2,-1/2)$-picture
degenerates simply into the vector emission vertex  
at $r\sim{1}$.Therefore the boundary contribution in ~\refs{\kleb}
had no peculiarities.
In the $(-3/2,-3/2)$ case the  situation is quite different.
Note that three additional poles depend exclusively on the
momentum of the incoming $(-3/2,-3/2)$ Ramond-Ramond particle
(and consequently, on the momentum of the 5-form branelike state).
 Such an unusual momentum dependence  may be
the consequence of the fact that the 
5-form vertex operator (3) is physical at some
discrete values of momentum only 
(such a behaviour would be reminiscent of that of discrete
states in two-dimensional gravity)
Another difference with the scattering amplitude 
in the $(-1/2,-1/2)$-picture is that in case of the D-instanton
background ($q=-1$)
the $(-3/2,-3/2)$ scattering amplitude
 does $not$ have  the field-theoretic structure observed in 
~\refs{\kleb}, on the contrary
due to the presence of the 5-form state the amplitude diverges
(since $k_1^{||}=0$ for the D-instanton background).
 As for the two-form branelike vertex operator 
$\sim{e^{-2\phi}\psi_m\psi_n}e^{ikX}$,
there is also a peculiarity at $k=0$.
Namely, at any nonzero  value of the momentum $k$ the two-form vertex
(3) may be decomposed into a sum of a 
fermionic part of a vector vertex operator in the $-2$-picture 
plus BRST-trivial part:
\eqn\grav{\eqalign{e^{-2\phi}\psi_m\psi_n{e^{ikX}}=A_{mn}+B_{mn}\cr
A_{mn}={1\over{k^2}}e^{-2\phi}(k\psi)(k_m\psi_n-k_n\psi_m)\cr
B_{mn}=e^{-2\phi}\psi_m\psi_n{e^{ikX}}-A_{mn}}}
To show that  one has to consider the correlation function
 of $e^{-2\phi}\psi_m\psi_n{e^{ikX}}$
with two other vector emission vertices in the $0$-picture on the sphere.
It is then straightforward to see that
\eqn\grav{\eqalign{\langle e^{-2\phi}
\psi_m\psi_n{e^{ikX}}V^0(p_1,w_1)V^0(p_2,w_2)\rangle=
\langle{A_{mn}}V^0(p_1,w_1)V^0(p_2,w_2)\rangle\cr 
\langle B_{mn}V^0(p_1,w_1)V^0(p_2,w_2)\rangle=0}}
where $V^0(p,w)=e_m(p)(\partial{X^m}+i(p\psi)\psi^m)e^{ipX}$.
Therefore at a nonzero momentum the two-form vertex operator (3)
can be expressed in terms of the picture $-2$ vector emission vertex,
however at $k=0$  such a decomposition is no longer possible and
the two-form becomes a new massless state.
\vfill\eject
\centerline{\bf Conclusion}

One of the claims of this work is that the superstring
action on the $AdS_5\times{S^5}$ can be viewed, from the point of view
of NSR formalism, as a sigma-model on the $flat$ spacetime;
introducing the background terms 
- a 5-form and a 3-form branelike terms is equivalent to
curving the SO(1,9) flat space-time to obtain the AdS background.
The $SO(1,3)\times{SO(6)}$ symmetry  of 
superstring theory on $AdS_5\times{S^5}$
space is manifest in the energy-momentum tensor (10).
Therefore we hope that exploring the dynamics of superstring theory
on $AdS_5\times{S^5}$ may be analogous to studying the correlation functions
of branelike vertex operators on the sphere.
As for the action corresponding to the stress-energy tensor (10)
 one should also add  terms corresponding to the Ramond-Ramond
background, corresponding to non-vanishing RR scalar field
in the near-core D-instanton background ~\refs{\gibb}
 These Ramond-Ramond terms should
correspond to the WZ term in  the action (4) (which does not
contribute to the stress- energy tensor).
The NSR action corresponding to the stress-energy tensor (10)
should, up to yet unknown RR part, look like
\eqn\grav{\eqalign{S_{{NSR}}=\int{d^2}z
\lbrack{1\over2}\partial{X^m}\bar\partial{X_m}
+\lbrace{1\over2}
:(\Gamma+\Gamma\bar\Gamma)\psi^{m}\bar\partial\psi_{m}:\cr+
\epsilon^{p_1p_2p_3p_4}\lbrack{e^\phi}\psi_{p_1}\psi_{p_2}
\psi_{p_3}\psi_{p_4}\psi_t\bar\partial{y^t}+\partial(e^\phi\psi_{p_1}
\psi_{p_2}\psi_{p_3})\bar\partial{{\tilde{x}}_{p_4}}\rbrack+ghosts
+RR-terms\rbrack}}
Note that since the 5-form branelike state has vanishing s-matrix
elements among elementary string states
(see also the discussion in the Appendix), adding branelike terms to
the NSR action in flat space-time does not 
affect perturbative string amplitudes and
does not create problems with unitarity.
Because of the ghost anomaly cancelation
condition adding the 5-form term to the action would not affect
the 2-point amplitude of ~\refs{\kleb} where the RR vertex operators
are taken in the standard picture.
Another crucial problem is related to the fact that
the branelike terms in the NSR stress-energy tensor (10) (as well as
possible Ramond-Ramond terms) have essentially nonzero
ghost number which cannot be removed by picture-changing transformation.
Therefore in order to explore the properties of the sigma-model
with the stress-energy tensor (10)
it is crucial to introduce some function of picture-changing operators
in the measure of functional integration in order to insure the
correct ghost number balance (that would cancel the ghost number anomaly).
The proper choice of such a measure deformation is an open question.
We hope that the conditions of the worldsheet conformal invariance
(i.e. vanishing of the 1-loop beta-function) which should be related
to requiring the functional with the action corresponding to the 
energy-momentum tensor (10)
 to satisfy the large N loop equation,
as well as  conditions for the amplitude factorization
will be sufficient to determine the correct deformation of the measure.
\centerline{\bf Acknowledgements}
It is a pleasure to thank  M.Douglas, N.Hambli, A.M.Polyakov and
S.Shatashvili
 for very important suggestions and discussions.
The author would like to express his deep gratitude to the European
Postdoctoral Institute (EPDI) for the support during the years
of 1996-1998 and to the Institut des Hautes Etudes Scientifiques
(IHES) for the hospitality during  the celebration of the 40th anniversary
of the IHES where I  elaborated on basic ideas of this work.
The author gratefully acknowledges the hospitality and support of
the Abdus Salam International Centre for Theoretical Physics
(ICTP) 
in Trieste and especially thanks S.Randjbar-Daemi.
 I'm also thankful to the physics department
of the University of Rome 2 ``Tor Vergata''
for the hospitality during the academic
year of 1997/98
(in particular it is a pleasure to thank M.Bianchi and A.Sagnotti).
\vfill\eject

\centerline{\bf Appendix}

We would like to comment briefly on the correlation
functions of the 5-form branelike state (3) with elementary
string states to demonstrate the vanishing of these correlators.
The vanishing of these correlation functions allows one to add
the branelike terms to the flat NSR action without creating
problems with the unitarity.  
Let us consider first the correlator of the 5-form
state with the vector emission vertices on the sphere.
The vector emission vertices in various pictures are given by:
\eqn\grav{\eqalign{V^{(0)}=e_m(k)(\partial{X^m}+i(k\psi)\psi^m)e^{ikX}\cr
V^{(-1)}(k)=e_m(k)e^{-\phi}\psi^m{e^{ikX}}\cr
V^{(-2)}(k)=e^{-2\phi}V^{(0)}(k)}}
On the sphere the total ghost number of correlators must be
equal to $-2$ to compensate the ghost anomaly.
First of all, it is clear that the 3-point function
and the 4-point function vanish:
\eqn\lowen{\langle V^{(-2)}(k_1)V^{(-1)}(k_2)e^\phi\psi_{m_1}...\psi_{m_5}
\rangle=
\langle V^{(-1)}(k_1)V^{(-1)}(k_2)V^{(-1)}(k_3)
e^\phi\psi_{m_1}...\psi_{m_5}\rangle\equiv
{0}}
simply because in both cases the total number of worldsheet
fermions $\psi$ in the vector vertices (taken together) is
less than 5 - so that there are some $\psi$'s left in the five-form vertex
operator with no partner to contract with.
The $n>4$-point correlation functions vanish also.
For the sake of brevity, let us demonstrate
it on the example of the 5-point function, for other $n$-point
functions the argument
remains the same.
Let us introduce the notation
$V_5=e^\phi\psi_{m_1}...\psi_{m_5}B^{{m_1}...{m_5}}$ where B is some rank 5
tensor which, without a loss of generality, may be assumed antisymmetric.
The 5-point function is given by:
\eqn\lowen{W_5=\langle V^{(0)}(k_1)V^{(-1)}(k_2)V^{(-1)}(k_3)
V^{(-1)}(k_4)V_5\rangle}
Clearly, it is only the fermionic part of the vertex $V^(0)$ that
could, in principle, lead to the non-vanishing contribution.
Consider two arbitrary fermions (say, $(k_1\psi)$ and $e(k_2)\psi$)
of the vector vertices
that couple with some 2 out of 5 fermions of $V_5$, say,
$\psi^{m_1}$ and $\psi^{m_2}$. The corresponding factor 
in the correlator would be given by
$k_1^{m_1}e^{m_2}(k_2)+k_1^{m_2}e^{m_1}(k_2)$.
But this factor has to be contracted with the tensor $B$ which is
antisymmetric in the indices $m_1$,$m_2$.Therefore the total
expression vanishes.
As for the correlation functions of $V^{(5)}$ with Ramond
vertices in the $standard$ $-1/2$-picture one can see that,
at least up to 6-point functions all the correlators vanish
(we would like to remind that the Ramond vertex operator
in the  $-3/2$-picture is $not$ the picture-changed version
of the standard picture $-1/2$ fermionic vertex.Rather, it is
a different physical state which $does$ have non-vanishing
matrix elements with $V_5$, as it is follows from
our calculation in the second section of this paper).
 As for the standard Ramond vertices,
all the correlation functions of the type
$\langle V^{(-1/2)}(k_1)...V^{(-1/2)}(k_{n-1})V_5\rangle$
 vanish on the sphere
at least for $n\leq{6}$
due to the ghost anomaly cancelation condition -
since the picture-changing factor  $f(\Gamma)$ 
(arising in the measure of the functional integral as a result 
of integration over supermoduli)
which regulates the ghost balance in the correlators, may only
increase ghost numbers but not reduce them.
For $n>6$, however, the proof of vanishing of   correlation functions
of the branelike vertex $V_5$ with standard Ramond states
 needs additional explicit computations.

\listrefs
  
\end